\documentstyle[pra,aps,amssymb,epsfig,multicol]{revtex}
\begin{document}
%
%
\title{Generic model of an atom laser}
\author{B.~Kneer,${}^{1,}$\cite{E.Mail}
T.~Wong,${}^2$ K.~Vogel,${}^1$ W.~P.\ Schleich,${}^1$ and D.~F.\ Walls${}^2$}
\address{${}^1$Abteilung f\"{u}r Quantenphysik, Universit\"{a}t Ulm, 
D-89069 Ulm, Germany\\
${}^2$Department of Physics, University of Auckland, Private Bag 92019,
Auckland, New Zealand}
\date{Submitted to Physical Review A: June 18, 1998;
revised manuscript submitted: August 20, 1998}
\maketitle
\begin{abstract}
We present a generic model of an atom laser by including a pump and loss
term in the Gross-Pitaevskii equation. 
We show that there exists a threshold for the pump above which the 
mean matter field assumes a non-vanishing value in steady-state.
We study the transient regime of this atom laser and find oscillations 
around the stationary solution even in the presence of a loss term. 
These oscillations are damped away when we introduce a 
position dependent loss term.
For this case we present a modified
Thomas-Fermi solution that takes into account the pump and loss. 
Our generic model of an atom laser is analogous to the
semi-classical theory of the laser.
\end{abstract}
\draft 
%
%
%
\begin{multicols}{2}
\narrowtext
\section{Introduction}
With the recent experiments on Bose-Einstein condensation 
\cite{BEC1,BEC2,BEC3,BEC4,BEC5,BEC6,BEC7,BEC8,BEC9,BEC10,BEC11,BEC12,reviews}
an atom laser, that is a device which produces an intense coherent 
beam of atoms by a stimulated process \cite{ketterle,wiseman}, has 
become feasible. Already, the MIT group has realized a pulsed 
atom laser \cite{mewes} and has provided evidence for the
process of coherent matter-wave amplification in the formation of
a Bose condensate \cite{miesner}. 

How to describe the matter field of an atom laser?
Rate equations are simple. But they cannot answer this question
since they do not contain any coherence.
In contrast a microscopic and fully quantum mechanical treatment 
can answer the question but is not easy to handle. 
What we need is a theory that includes coherence but is still 
simple. In the present paper we develop
such a theory---a generic model of the atom laser. 

The optical laser has three essential ingredients:
(i) a resonator for the electromagnetic field,
(ii) an atomic medium, and
(iii) an excitation mechanism for the atoms.
We start the laser cycle by preparing the atoms in an excited state 
using this excitation mechanism, which in general is incoherent.
The radiation emitted by the atom amplifies the electromagnetic field
in a mode of the resonator. To be efficient the frequency of the
mode has to match appropriately the frequency of the transition.
The boundary conditions set by the resonator determine
via the Helmholtz equation the spatial part of the mode function. 
The ultimate goal is to transfer the excitation of the atom into
a macroscopic excitation of the field mode. In this way we transfer 
the energy used to excite the atom via the gain medium 
into coherent excitation of the field mode.
In order to make use of the radiation we have to couple it out 
of the resonator. We compensate for this loss by
continuously re-pumping the medium. 

The goal of an atom laser is completely analogous:
We want to create a macroscopic coherent excitation of a
mode in a resonator for atoms. Hence, in an atom laser the atoms 
play the role of the field excitation of the optical laser.
Since atoms cannot be created or
annihilated the means to achieve lasing are different.
Indeed, in an atom laser there is no ``real'' laser medium: 
It is the same atom that goes through the laser process.
We only manipulate the internal degrees of freedom and 
center-of-mass motion of the atoms. In particular we want to
force their center-of-mass motion into a specific 
quantum state of the resonator. 
The resonator for the atoms is a binding potential such as 
provided by a trap. The spatial part of the mode function of this 
atomic resonator follows from the time independent Schr\"odinger 
equation \cite{dipdip}. Moreover, we focus on the ground state of 
the trap. As in the optical laser we want a macroscopic excitation 
of this mode, that is we strive to have as many atoms as possible
in one quantum state. This is the phenomenon of Bose-Einstein 
condensation. As in the optical laser we need to couple the atomic 
wave out of the atomic resonator. In order to have a continuous 
wave (cw) atom laser, we have to continuously feed in more atoms.

There exist two different approaches
towards a theoretical description of a cw atom laser:
The first one relies on rate equations 
\cite{rateoptical1,rateoptical2,rateoptical3} whereas the
second one derives a quantum mechanical master equation 
\cite{masteroptical1,masteroptical2,masterdipdip,masterevap1,%
masterevap2,masterdissoc}. 
In the present paper we suggest a third approach which makes
heavily use of the close analogy between an atom laser and an 
optical laser. In the latter case it 
turned out that a classical treatment of the electromagnetic 
field \cite{haken,lamb} was sufficient to describe many features 
of the laser. Can we therefore devise a semi-classical theory of 
the atom laser?

The semi-classical laser theory replaces the electromagnetic 
field operator $\hat{E}({\bf r},t)$ for the field inside the 
laser cavity by the expectation value 
${\cal E}({\bf r},t)\equiv\langle\hat{E}({\bf r},t)\rangle$. 
The equation of motion for ${\cal E}$ is the wave equation of 
Maxwell's electrodynamics driven by the polarization of the laser 
medium. An additional term introduced phenomenologically takes into 
account the loss of the cavity. The polarization of the laser medium 
follows from a microscopic, quantum mechanical description of the 
internal structure of the atoms. Quantum mechanics rules the atoms 
whereas classical Maxwell's wave theory determines the electromagnetic 
field. These are the essential ideas of semi-classical laser theory.

The semi-classical laser equations do not prefer any particular phase.
Nevertheless by choosing an arbitrary phase we can describe many properties
of the electromagnetic field. Furthermore, we have to start with a 
non-vanishing seed field in order to obtain a non-vanishing
solution for the electromagnetic field with a fixed phase.

In our model of an atom laser we replace the matter-wave field
represented by a field operator $\hat{\Psi}({\bf r},t)$ by a scalar 
mean field $\psi({\bf r},t)\equiv\langle\hat{\Psi}({\bf r},t)\rangle$.
This is analogous to replacing the field operator $\hat{E}$ by its 
expectation value ${\cal E}$ in semi-classical laser theory. 
The well-known Gross-Pitaevskii equation \cite{GPE1,GPE2} plays now 
the role of Maxwell's wave equation. It defines the equation of motion
for $\psi({\bf r},t)$. 
Similar to the semi-classical theory of the optical laser we have to
break the symmetry of the equation of motion for $\psi$ in order to
have a non-vanishing value for $\psi$. For massive particles this is
more problematic than for photons since they cannot be created or
annihilated, and for all quantum states with fixed particle number
we have $\langle\hat{\Psi}\rangle =0$. Nevertheless the concept of
spontaneously broken symmetry turned out to be very useful to describe 
properties of a condensate, in particular interference effects. For more
detailed discussions see Refs.\ \cite{wiseman,moelmer}.
In contrast to the driven electromagnetic wave
equation the Gross-Pitaevskii equation does not contain a gain term 
analogous to the polarization. This reflects the fact that in 
Bose-Einstein condensation there is no ``medium'' in the trap. We 
therefore add a phenomenological pump term. Moreover,
as in the electromagnetic case we have to add a loss term. 

Despite the similarity there is a fundamental difference 
in the two equations of motion.
In the absence of a medium Maxwell's wave equation is linear. 
In contrast the Gross-Pitaevskii equation is non-linear.
This is a manifestation of the interaction of atoms.

The crucial part of any laser is the stimulated amplification process. 
Different mechanisms for matter-wave amplification have been suggested
and discussed: optical cooling 
\cite{rateoptical1,rateoptical2,rateoptical3,masteroptical1,masteroptical2},
elastic collisions by evaporative cooling 
\cite{masterevap1,masterevap2,gardiner}, 
dissociation of molecules \cite{masterdissoc},
and cooling by a thermal reservoir \cite{scully}.
For different schemes of pumping a condensate we refer to Refs.\ 
\cite{pump1,pump2,pump3}.
Our model does not rely on a specific mechanism, but 
can be adapted to any mechanism where Bose enhancement is present. 

The paper is organized as follows:
In Sec.\ \ref{sec:model}, we generalize the Gross-Pitaevskii equation 
by including gain and loss terms. One important loss is due to coupling 
the atom wave out of the resonator. However, similar to semi-classical 
laser theory where usually the field inside the laser cavity is 
investigated, we restrict ourselves to the matter-wave field inside 
the resonator. We therefore do not go into details of an output coupler
\cite{output1,output2,output3,output4,output5,output6}. We present 
the stationary solutions of our equations and perform a stability 
analysis. We find a threshold behavior similar to the optical laser. 
Moreover, we calculate the time dependent
solution and show how it converges to a quasi-stationary solution. 
In general, the quasi-stationary solution we find does not coincide 
with the time independent solution but shows some oscillatory behavior
around it. These oscillations disappear when we modify our 
equations in Sec.\ \ref{sec:spaceloss} by introducing a 
space dependent loss. This loss can be thought of as a consequence 
of collisions between condensed and un-condensed atoms at the edges of 
the condensate. For this improved model we present a modified 
Thomas-Fermi approximation for the stationary solution. 
Section \ref{sec:concludere} summarizes our results.
\section{Elementary model}
\label{sec:model}
In this section we summarize our generic model of an atom laser.
\subsection{Formulation of the model}
\label{subsec:formulation}
In particular, we add the phenomenological gain and loss terms
to the Gross-Pitaevskii equation (GPE). In this way we couple the
GPE to an equation governing the number of un-condensed atoms.

The Gross-Pitaevskii equation (GPE) for the mean field $\psi=\psi({\bf r})$ 
of the condensed atoms of mass $m$ in a trap potential $V({\bf r})$ 
reads \cite{BEC-GPE1,BEC-GPE2} 
\begin{equation}  \label{GPeq}
i\hbar \frac{\partial \psi}{\partial t} 
=-\frac{\hbar^2}{2m}\Delta\psi+V({\bf r})\psi+U_0|\psi|^2\psi.
\end{equation}
The non-linear term $U_0|\psi|^2\psi$ with $U_0=4\pi\hbar^2 a_s/m$ 
takes into account two-particle interactions where $a_s$ denotes
the s-wave scattering length. 

The GPE has been very successful in describing the properties of 
Bose-Einstein condensates. However, in the present form it cannot 
describe the growth of or the loss of atoms out of a condensate. 
Indeed, the GPE keeps the number of atoms 
\begin{equation}\label{DefNg}
N_c \equiv\int |\psi({\bf r})|^2 \, d^3r
\end{equation}
constant. 

In order to overcome this problem
we introduce two additional terms in the GPE: a loss term and
a gain term. The loss term 
\begin{equation}  \label{lossterm}
H_{{\rm loss}} \psi \equiv -\frac{i\hbar}{2}\gamma_c \psi
\end{equation}
leads to an exponential decay of the number of atoms in the 
condensate. In contrast the gain term 
\begin{equation}  \label{gainterm}
H_{{\rm gain}} \psi \equiv \frac{i\hbar}{2}\Gamma N_u \psi 
\end{equation}
leads to an increase of atoms in the condensate. Here $N_u$ is the 
number of atoms outside the condensate, that is, the un-condensed atoms, 
and $\Gamma$ is the rate for the transition of these atoms into the 
condensate. We regard this as a generic 
pump mechanism \cite{mechanism} of an atom laser since it contains 
already the Bose enhancement as we will see later. 

When we add the loss term (\ref{lossterm}) and 
the gain term (\ref{gainterm}) to the GPE (\ref{GPeq}), 
we arrive at the generalized GPE 
\begin{eqnarray}  \label{psi}
i\hbar \frac{\partial \psi}{\partial t} 
&=&-\frac{\hbar^2}{2m}\Delta\psi+V({\bf r})\psi +U_0|\psi|^2\psi 
\nonumber\\
&-&\frac{i}{2}\hbar\gamma_c\psi+\frac{i}{2}\hbar\Gamma N_u\psi.
\end{eqnarray}
For the number of un-condensed atoms $N_u$ we assume the rate equation 
\begin{equation}  \label{Ne}
\dot{N}_u=R_u-\gamma_u N_u-\Gamma N_c N_u.
\end{equation}
The first term reflects the fact that we generate with a rate $R_u$
atoms in an un-condensed state from an infinite reservoir of atoms.
The term $-\gamma_u N_u$ takes into account that atoms can escape from 
our system without being trapped in the condensed state. 
The last term describes the transition of the atoms into the condensate.
Since it is proportional to the number of atoms $N_c$ in the condensate,
it contains the Bose-enhancement factor. 

In this way we have coupled the generalized GPE governing the
condensate to the rate equation 
governing the number of un-condensed atoms.
These two equations (\ref{psi}) and (\ref{Ne}) are the foundations 
of our model.

The number $N_c$ of condensed atoms follows from the generalized 
GPE (\ref{psi}) with the help of the definition, Eq.\ (\ref{DefNg}).
From Eq.\ (\ref{psi}) we obtain the rate equation
\begin{equation}  \label{Ng}
\dot{N}_c=\Gamma N_u N_c -\gamma_c N_c
\end{equation}
for the number of atoms in the condensate.

We conclude this section by noting that rate
equations similar to Eqs.\ (\ref{Ne}) and (\ref{Ng}) have already been
discussed in the literature \cite{rateoptical1,rateoptical2}.
However, in the present paper we replace the rate equation
(\ref{Ng}) by the generalized GPE (\ref{psi}). This equation obviously
contains the rate equation but in addition the coherence.
\subsection{Stationary solutions and stability analysis}
\label{subsec:stab}
Before we turn to the stationary solution of the generalized GPE,
Eq.\ (\ref{psi}), we first
discuss the stationary solutions of the rate equations (\ref{Ne}) 
and (\ref{Ng}).
\subsubsection{Rate equations}
\label{subsec:rate}
The rate equations (\ref{Ne}) 
and (\ref{Ng}) have two possible stationary solutions:

The solution
\begin{eqnarray}
N_c^{(s)} &\equiv &0, \nonumber\\
N_u^{(s)} &=& R_u/\gamma_u 
\label{ratestatsol1}
\end{eqnarray}
is reminiscent of the optical laser below threshold where
the intensity of the laser vanishes.

The other stationary solution 
\begin{eqnarray}
N_c^{(s)}&=&\frac{R_u}{\gamma_c}-\frac{\gamma_u}{\Gamma}, \nonumber \\
N_u^{(s)}&=&\frac{\gamma_c}{\Gamma}
\label{ratestatsol2}
\end{eqnarray}
corresponds to the laser above threshold. 

We now perform a stability analysis of these solutions.
For this purpose we 
introduce small deviations 
\begin{eqnarray}
n_u(t) &\equiv& N_u(t) - N_u^{(s)},  \nonumber \\
n_c(t) &\equiv& N_c(t) - N_c^{(s)},
\end{eqnarray}
and arrive at the linearized equations 
\begin{eqnarray}
\dot{n}_u &=& -(\gamma_u +\Gamma N_c^{(s)}) n_u - \Gamma N_u^{(s)} n_c , 
\nonumber \\
\dot{n}_c &=& -(\gamma_c - \Gamma N_u^{(s)}) n_c + \Gamma N_c^{(s)} n_u .
\label{linearized}
\end{eqnarray}
A stability analysis of these equations shows that the stationary 
solution in Eqs.\ (\ref{ratestatsol1}) is stable for
\begin{equation}\label{threshold}
R_u < R^{th}\equiv\frac{\gamma_c\gamma_u}{\Gamma}. 
\end{equation}
Likewise, our stability analysis shows that Eqs.\ (\ref{ratestatsol2})
are a stable stationary solution for 
\begin{equation}
R_u > \frac{\gamma_c\gamma_u}{\Gamma}= R^{th}.  
\end{equation}
Therefore, there exists a threshold $R^{th}$.
When the pump rate $R_u$ is below the threshold,
the number $N_c$ of atoms in the condensate vanishes. When it is above, 
it is non-vanishing. In that case the steady-state number of
atoms in the condensate grows linearly with the pump rate $R_u$.
\subsubsection{Generalized GPE}
We now turn to the discussion of the stationary 
solutions of our generic model of the atom laser, that is of
the generalized GPE coupled to the rate equation for $N_u$.

The stable stationary 
solution of Eq.\ (\ref{psi}) corresponding to a vanishing number of
atoms in the condensate is
\begin{equation}
\psi^{(s)} \equiv 0. 
\end{equation}
This is the stationary solution of the mean field below threshold.

Above threshold
the stationary solution $\psi^{(s)}$ of the generalized GPE (\ref{psi})
is identical to the stationary solution 
of the conventional GPE. Indeed, when we substitute the ansatz
$\psi=\exp(-i\mu t/\hbar) \psi^{(s)}$ into Eq.\ (\ref{psi})
and note that the gain and 
loss terms cancel each other as a result of the stationary solution, 
Eq.\ (\ref{ratestatsol2}), we arrive at 
\begin{equation}  \label{GPeqtimeindep}
\mu\psi^{(s)} =-\frac{\hbar^2}{2m}\Delta\psi^{(s)}+
V({\bf r})\psi^{(s)}+U_0|\psi^{(s)}|^2\psi^{(s)}.
\end{equation}
Here, $\mu$ denotes the chemical potential. 

A stability analysis of this
equation leads to collective excitations \cite{excitations} which
in general cannot be treated analytically. 
For such an analysis in the case of a one-dimensional harmonic
oscillator we refer to the Appendix.
\subsection{Transient behavior}
\label{subsec:num}
In the preceding section we have discussed the steady-state 
solutions of both the rate equations (\ref{Ne}) and (\ref{Ng})
and the matter-field equations (\ref{psi}) and (\ref{Ne}) of the 
atom laser. In the present section we address the question 
if and how the matter field evolves 
into a stationary state from an initial condition. 
Here we focus on the case above threshold.
Since the equations are non-linear we solve them numerically
for the case of an one-dimensional harmonic oscillator potential
\begin{equation}\label{harmpot}
V(x)=\frac{1}{2} m \omega^2 x^2. 
\end{equation}
This one-dimensional trap potential
already shows the essential features of the atom laser.

We therefore analyze the one-dimensional generalized GPE
\begin{eqnarray}  \label{psi1D}
i\hbar \frac{\partial}{\partial t}\psi(x,t) 
&=&\left[-\frac{\hbar^2}{2m}\frac{\partial^2}{\partial x^2} 
+\frac{1}{2}m \omega^2 x^2 +U_x|\psi(x,t)|^2 \right.  
\nonumber \\
&&-\left.\frac{i}{2}\hbar\gamma_c
+\frac{i}{2}\hbar\Gamma N_u\right]\psi(x,t),
\end{eqnarray}
describing the matter field, coupled to the equation 
\begin{equation}\label{Ne1D}
\dot{N}_u=R_u-\gamma_u N_u-\Gamma N_c N_u
\end{equation}
for the number of un-condensed atoms. We use the
split-operator method \cite{split1,split2} to solve the generalized 
GPE. This technique has already been used
successfully for the ordinary GPE \cite{munich}.
Moreover, we supplement this numerical analysis by an analytical
treatment of the solution in the long-time limit where we apply
the method of Ref.\ \cite{coll2} to one dimension.
For the details of the analytical approach we refer to the
Appendix.
\subsubsection{Initial conditions and parameters}
The semi-classical theory of the optical laser cannot explain the
initial start-up of the laser since it does not contain
spontaneous emission: 
A non-vanishing seed electromagnetic field starts the laser. 
Similarly, in our model of the atom laser we start from an
initial condition for the mean matter field $\psi(x,t=0)$ that
is non-zero. This is the seed field for our atom laser.

Let us illustrate this by considering for the moment
the ``natural'' initial condition $\psi(x,t=0)=0$ and $N_u(0)=0$. 
This implies $N_c(0)=0$. We therefore start 
without any atoms in the system. Moreover, we consider a pump rate
$R_u$ above threshold. This choice of initial conditions leads to 
the unstable solution $N_c=0$ and $N_u=R_u/\gamma_u $. Indeed, any small 
perturbation in $N_c$ leads to a completely different behavior and 
$N_c$ and $N_u$ approach the stable solutions 
$N_c^{(s)}=R_u/\gamma_c-\gamma_u/\Gamma $ and 
$N_u^{(s)}=\gamma_c/\Gamma $, as
discussed in Sec.\ \ref{subsec:stab}. Therefore, we use the different
initial condition $N_c(0)\ll 1$ for our numerical 
simulations \cite{discnumrem}. We keep the condition $N_u(0)=0$.

For our numerical calculations we take the parameter 
$U_x/(\hbar\omega a) \cong 0.008$,
where $a\equiv\sqrt{\hbar/(m\omega)}$ is the width of the 
ground state of the harmonic oscillator. 
As in conventional laser theory \cite{haken,lamb}, where one usually
has a high-Q cavity, we require $\gamma_c \ll \omega$. 
Since in the steady state we do not want to have too many atoms in the 
non-condensate part, we require $\gamma_c\lesssim\Gamma$, 
as suggested by Eq.\ (\ref{ratestatsol2}).
In order to have a reasonably small threshold value we choose
$\gamma_u$ appropriately, as indicated
by Eq.\ (\ref{threshold}).
We also need to choose $R_u$ such that we have a sufficiently 
large number of atoms in the
condensate, that is, $N_c^{(s)} = R_u/\gamma_c-\gamma_u/\Gamma \gg 1$.
\subsubsection{Average properties of the matter field}
In Fig.~\ref{figure1}(a) we show the time dependent solutions $N_c(t)$
and $N_u(t)$ of the rate equations (\ref{Ne}) and (\ref{Ng}). For
very small times the atoms accumulate in the un-condensed phase: Due 
to the small number of atoms in the condensate the Bose enhancement is 
not yet effective and they cannot make a transition into the condensate.
As soon as we have a significant number 
of atoms in the condensate, $N_u$ rapidly approaches its stationary 
value $N_u^{(s)}=\gamma_c/\Gamma$. Additional atoms then essentially end 
up in the condensate where the number of atoms slowly approaches its 
stationary value $N_c^{(s)}=R_u/\gamma_c - \gamma_u/\Gamma$.

For the numerical analysis of the Eqs.\ (\ref{psi1D}) and (\ref{Ne1D}) 
defining our generic model of the atom laser, we use 
the same parameters. For the initial condition $\psi(x,t=0)$ of the 
generalized GPE we use the quantum mechanical 
ground state of the trap, normalized in such a way that 
$N_c(0) = \int |\psi(x,t=0)|^2 \,dx$. 
In order to get some feeling for the accuracy of our split-operator 
technique for the generalized GPE we calculate 
$N_c(t) = \int |\psi(x,t)|^2 \,dx$ which according to 
Sec.\ \ref{subsec:formulation} has to coincide with the solution of 
the rate equations. 

We gain insight into the time dependence of 
$|\psi(x,t)|^2$ by calculating its first moment 
\begin{equation}\label{firstmom}
\overline{x(t)} = \frac{\int x |\psi(x,t)|^2 \,dx}
           {\int |\psi(x,t)|^2 \,dx}
\end{equation}
and second moment \cite{noteqmexp}
\begin{equation}\label{secondmom}
\overline{x^2(t)} = \frac{\int x^2 |\psi(x,t)|^2 \,dx}
           {\int |\psi(x,t)|^2 \,dx} .
\end{equation}

We note that the symmetry of the trap and the generalized GPE ensure 
that the symmetry of our initial condition is preserved, that is 
\begin{equation}
\overline{x(t)}=0. 
\end{equation}
%
\centerline{\epsfig{file=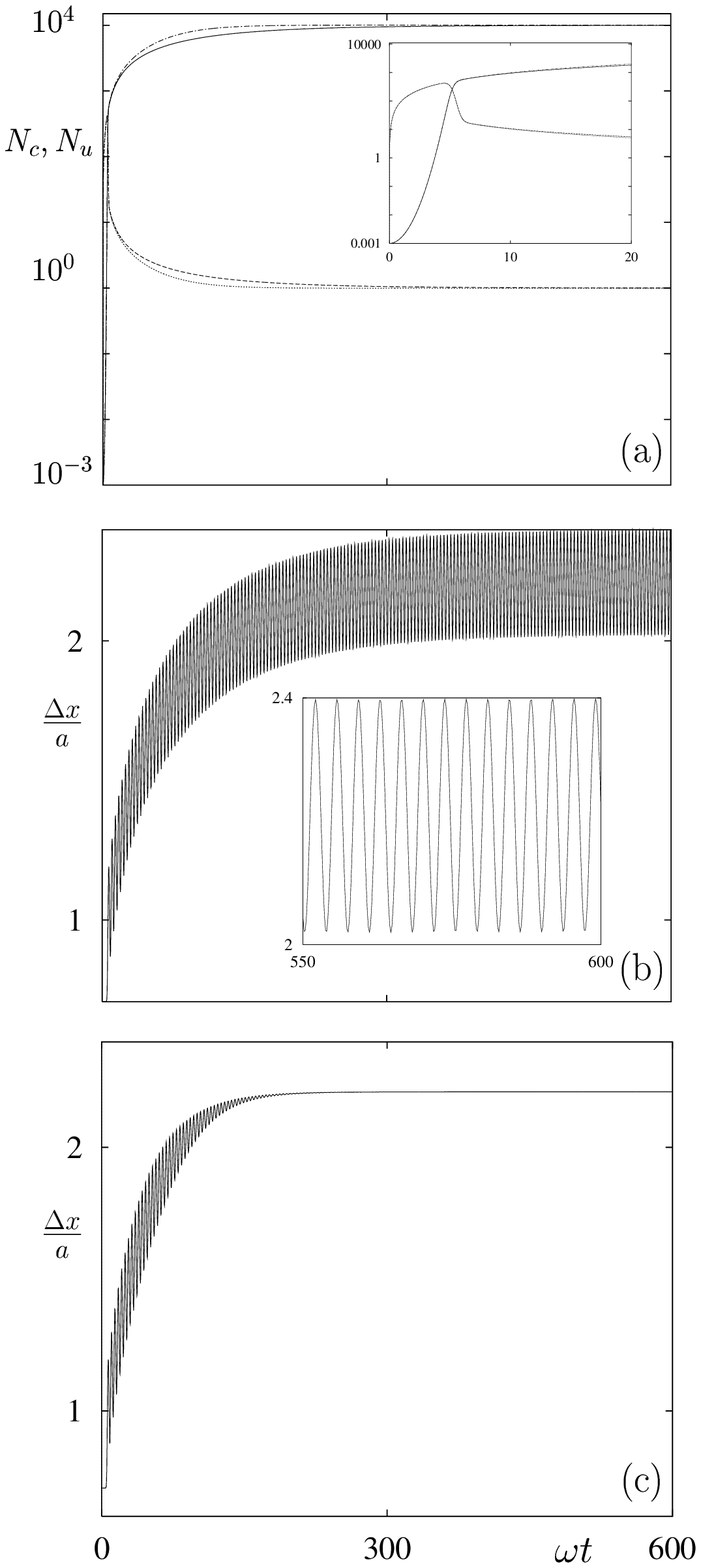,width=8.5cm}}   
\begin{figure}[h] 
\caption{Generic model of an atom laser: transient regime.
(a) The number $N_c$ of atoms in the condensate 
for the spatially independent (solid line) and spatially dependent
(dash-dotted line) loss,
and the number of un-condensed atoms $N_u$ for the spatially independent 
(dashed line) and spatially dependent (dotted line) loss
approach their steady-state values $10^4$ and 1, respectively.
The scale on the vertical axis is logarithmic. The inset shows 
the early stage of the transient regime. Here, we cannot distinguish
the curves corresponding to the different spatial damping, since 
for small atom numbers the spatial dependence does not play a big role.
(b) The width $\Delta x/a$ of the mean field as a function
of time displays collective excitations with period 
$T=3.57/\omega$ as apparent from the inset. For the spatially 
homogeneous damping the oscillations are un-damped whereas
for the spatially dependent loss they are damped away 
as shown in (c). The parameters are $R_u/\omega =10^2$, 
$\Gamma/\omega =10^{-2}$, $\gamma_u/\omega =10^{-2}$,
$\gamma_c/\omega =10^{-2}$, 
and $U_x/(\hbar\omega a)\protect\cong 0.008 $.
As initial conditions we have used $N_u(0)=0$, $N_c(0)=10^{-3}$, and
for $\psi(x,t=0)$ we have used the quantum mechanical  
ground state of the trap, normalized in such a way that  
$\int |\psi(x,t=0)|^2 \,dx =N_c(0)$. 
For (c) we have used a spatially dependent loss $\gamma_c(x)$
described by Eq.\ (\ref{gaussianloss}) with 
$\gamma_c'/\omega=0.192$, 
$x_0/a =5$, and $\sigma/a=1$.
} 
\label{figure1} 
\end{figure} 
\noindent
Hence the width 
\begin{equation}
\Delta x(t)\equiv \sqrt{\overline{x^2} - \overline{x}^2} 
= \sqrt{\overline{x^2(t)}}
\end{equation}
of $|\psi(x,t)|^2$ is governed by the second moment, only.

In Fig.\ \ref{figure1}(b) we plot the width $\Delta x$. 
We note that $\Delta x$ increases as a function of time
and approaches a steady-state value for large times. Moreover,
it oscillates around this steady-state value. In the inset we magnify 
these oscillations. We note that the oscillations do not decay
\cite{reviv}.
Therefore $\Delta x$ does not approach a time independent value and
there is no steady state in a strict sense.
We hence refer to this solution as the quasi-stationary solution.

In the Appendix we derive an analytical expression
for the frequency $\Omega_2$ of these oscillations and find
$\Omega_2= \sqrt{3}\,\omega \cong 1.73\,\omega$.
From Fig.\ \ref{figure1}(b) we read off 
$\Omega_2 \cong 1.76\,\omega$ which is in good agreement with our 
prediction. 
\subsubsection{Matter field}
\begin{figure}[h] 
\centerline{\epsfig{file=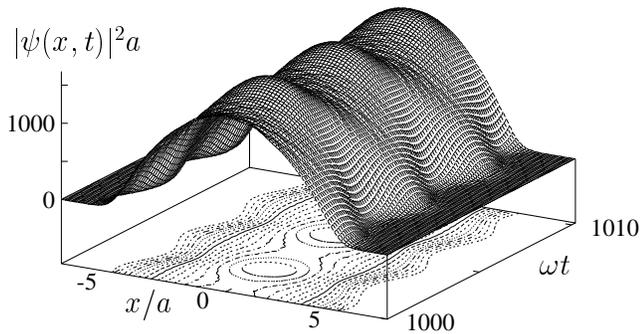,width=8.5cm}}  
\caption{Quasi-stationary solution $|\psi(x,t)|^2$ of our atom-laser 
equations for a time interval when the number of atoms in
the condensate has already reached its steady-state value.
We have used the same parameters as in Fig.\ \ref{figure1}.} 
\label{figure2} 
\end{figure}

The oscillations in the width of the distribution are a manifestation 
of collective excitations \cite{excitations} of the condensate. Indeed, the 
whole distribution oscillates as shown in Fig.\ \ref{figure2}.
Here we display the 
mean field for a time interval where the numbers of atoms in the
condensed state and the un-condensed state have already reached their
steady-state value. 

In order to study this oscillatory behavior in more detail,
we compare in Fig.\ \ref{figure3}(a) the solution 
of the time independent GPE (\ref{GPeqtimeindep}) in one dimension 
to the quasi-stationary solution of Eqs.\ (\ref{psi1D}) and (\ref{Ne1D}). 
Here we have depicted $|\psi(x,t)|^2$ for time moments where the 
width is 
\noindent
maximal, average, and minimal. Indeed, the 
solution of Eqs.\ (\ref{psi1D}) and (\ref{Ne1D}) oscillates 
around the solution of the time independent 
GPE \cite{statsols1,statsols2,statsols3}. 
\section{Improved model}
\label{sec:spaceloss}
The collective excitations discussed in the preceding section depend
on the initial condition $\psi(x,t=0)$. This is due to the fact that so far
there is no mode selection mechanism in our model which would favor one
stationary solution of the GPE over the others.
This fact is similar to the optical multi-mode laser
\cite{multimodelaser1,multimodelaser2}.
We can accomplish a single-mode atom laser 
when we allow for a space dependent loss. This is the topic of
the present section where we formulate an improved model 
for an atom laser.
\subsection{Formulation of the problem}
Real losses in a trapped condensate are spatially dependent: 
For example, a possible loss mechanism are collisions with un-condensed 
atoms which are preferably located at the edge of the condensate, see 
Refs.\ \cite{BEC1,BEC2,BEC3,BEC4,BEC5,BEC6,BEC7,BEC8,BEC9,BEC10,BEC11,BEC12}.
Further, atoms can be lost at the edge of the condensate because of
a finite trapping potential. Moreover, the goal of a matter-wave 
output-coupler is to create a directed coherent atomic beam. All of these 
reasons require a spatially dependent loss term. In contrast to this loss 
term we keep the pump term spatially independent. This reflects the
fact that we assume that the condensate is pumped from a cloud
of cold thermal atoms which is larger in size.

We therefore consider a modified generalized GPE
\begin{eqnarray}  \label{psiSL}
i\hbar \frac{\partial \psi}{\partial t} 
&=&-\frac{\hbar^2}{2m}\Delta\psi 
+V({\bf r})\psi +U_0|\psi|^2\psi  \nonumber \\
&-&\frac{i}{2}\hbar\gamma_c({\bf r})\psi+\frac{i}{2}\hbar\Gamma N_u\psi,
\end{eqnarray}
\begin{figure}[h] 
\centerline{\epsfig{file=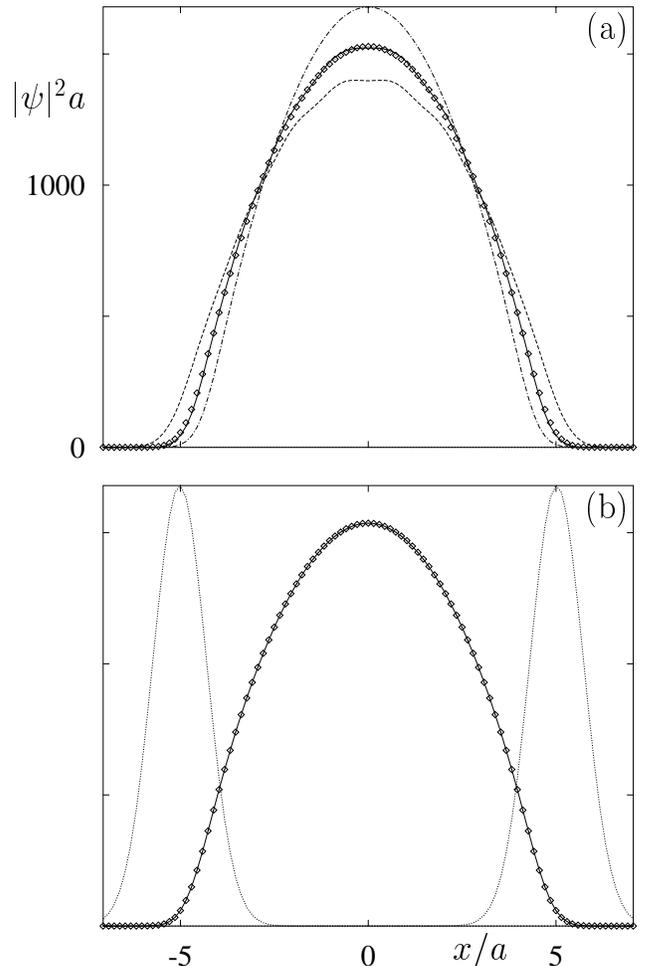,width=8.5cm}}  
\caption{Quasi-stationary and stationary solutions of our 
atom-laser equations. 
On the top (a) we compare the solution of the time independent 
Gross-Pitaevskii equation (diamonds) to the quasi-stationary 
solution of Eq.\ (\ref{psi1D}) for times where the width is 
maximal (dashed line), average (solid line), and minimal 
(dash-dotted line), respectively. These times correspond to 
$\omega t=1000.54$, $1001.45$, $1002.35$. 
On the bottom (b) we consider the case of spatially dependent loss. 
We depict (not to scale) the loss rate $\gamma_c(x)$ by a dotted line.
In this case collective excitations are damped out
and in steady-state the solutions of our atom laser equations 
(\ref{psiSL}) and (\ref{NeSL}) coincide with the time independent 
ground-state solution (diamonds) of Eqs.\ (\ref{ratesolsSL1g}) and 
(\ref{time_indep}). In both figures we have used the parameters 
of Fig.\ \ref{figure1}.} 
\label{figure3} 
\end{figure} 
\noindent
where the decay rate $\gamma_c({\bf r})$ is spatially dependent. 
Hence, the rate equation (\ref{Ng}) for the number $N_c$ of condensed
atoms now translates into
\begin{equation} \label{NgSL}
\dot{N}_c=\Gamma N_u N_c 
-\int\gamma_c({\bf r}) |\psi({\bf r},t)|^2 \,d^3r.
\end{equation}
Note that due to the space dependent loss rate we cannot express
the integral in Eq.\ (\ref{NgSL}) by $N_c$. Therefore we do not
have a rate equation for $N_c$ anymore.
However, the rate equation 
\begin{equation} \label{NeSL}
\dot{N}_u=R_u-\gamma_u N_u-\Gamma N_c N_u
\end{equation}
for the number $N_u$ of un-condensed atoms remains the same.

Equations (\ref{psiSL}) and (\ref{NeSL}) are the two fundamental
equations of our improved model of an atom laser.
\subsection{Stationary solution}
\label{subsec:statsol}
We now study the stationary solution of Eqs.\ (\ref{psiSL}) and (\ref{NeSL}). 
We first derive the steady-state expressions 
and then discuss the laser threshold in the
presence of a position dependent loss term. We conclude by introducing 
a modified Thomas-Fermi approximation.
\subsubsection{Matter field in steady-state}
When we follow an analysis similar to the one of Sec.\ \ref{subsec:stab}, 
we find two sets of stationary solutions of the Eqs.\ (\ref{psiSL}), 
(\ref{NgSL}), and (\ref{NeSL}).
The first set with 
\begin{eqnarray}  
N_u^{(s)} &=& \frac{R_u}{\gamma_u}, \nonumber\\
N_c^{(s)} &=& 0,\nonumber\\
\psi^{(s)} &=& 0
\label{ratesolsSL2}
\end{eqnarray}
corresponds to the solution below threshold.

The second set 
\begin{eqnarray}  
N_u^{(s)} &=& 
\frac{1}{\Gamma}\int\gamma_c({\bf r})|\psi^{(s)}_1({\bf r})|^2 \,d^3r 
=\frac{R_u}{\gamma_u+\Gamma N_c^{(s)}},
\label{ratesolsSL1e}\\
N_c^{(s)} &=& 
\frac{R_u}{\int\gamma_c({\bf r})|\psi^{(s)}_1({\bf r})|^2 \,d^3r}
-\frac{\gamma_u}{\Gamma}
\label{ratesolsSL1g}
\end{eqnarray}
corresponds to the solutions above threshold as we show now. Here, 
$\psi^{(s)}_1({\bf r})=[N_c^{(s)}]^{-1/2}\psi^{(s)}({\bf r})$ 
denotes the stationary solution of 
Eq.\ (\ref{psiSL}) normalized to unity. This solution is defined
by the time-independent modified generalized GPE
\begin{eqnarray}\label{time_indep}
\mu\psi^{(s)} 
&=&-\frac{\hbar^2}{2m}\Delta \psi^{(s)} +V\left( {\bf r}\right)
\psi^{(s)} +U_0 |\psi^{(s)}|^2\psi^{(s)}  \nonumber \\
&-&\frac{i}{2}\hbar\gamma_c\left( {\bf r}\right) \psi^{(s)} 
+\frac{i}{2}\hbar\Gamma N_u\psi^{(s)} ,
\end{eqnarray}
following from Eq.\ (\ref{psiSL}) using the ansatz
$\psi({\bf r},t) =\exp(-i\mu t/\hbar)\psi^{(s)}({\bf r})$ 
where $\mu$ denotes the ``chemical potential.''

These expressions for $N_u^{(s)}$ and $N_c^{(s)}$ are not explicit since 
they involve the stationary solution $\psi^{(s)}({\bf r})$ of 
Eq.\ (\ref{time_indep}). In Sec.\ \ref{subsec:modTFsol} we derive 
an approximate analytical expression and in Sec.\ \ref{subsec:exactsol} 
we find a fully numerical solution for $\psi^{(s)}({\bf r})$. 
\subsubsection{Lasing threshold}
A stability analysis of the above solutions amounts to calculating
the collective excitations of the modified generalized GPE (\ref{psiSL}).
This is only possible numerically. We therefore pursue a strategy
where we solve numerically for the full time dependence of
$\psi({\bf r},t)$. In order to gain some insight into the 
threshold condition we first discuss simple physical arguments.

These considerations rely on the fact that the number of atoms cannot 
be negative. With the help of Eq.\ (\ref{ratesolsSL1g}) we then 
determine the laser threshold 
\begin{equation}\label{SLthres0}
R^{th}
\equiv\frac{\gamma_u}{\Gamma}
\int\gamma_c({\bf r})|\psi^{(s)}_1({\bf r})|^2 \,d^3r
\end{equation}
by setting $N_c^{(s)}=0$. 
Close to threshold the number $N_c^{(s)}$ of atoms in the condensate
is close to zero.
We can therefore neglect the non-linear contribution in the modified
generalized GPE and arrive at a linear Schr\"odinger equation
with pump and loss. With a position dependence of the loss term
that favors the normalized ground-state energy solution $\phi_0({\bf r})$ 
of the linear Schr\"odinger equation, we can replace $\psi^{(s)}_1({\bf r})$ 
by $\phi_0({\bf r})$ in Eq.\ (\ref{SLthres0}) and arrive at
\begin{equation}\label{SLthres2}
R^{th}
=\frac{\gamma_u}{\Gamma}\int\gamma_c({\bf r})|\phi_0({\bf r})|^2 \,d^3r.
\end{equation}
When the loss shape is flat around the localization of
$\phi_0({\bf r})$, that is, around the
center ${\bf r}={\bf 0}$ of the trap, we can factor out
$\gamma_c({\bf 0})$ from the integral
in Eq.\ (\ref{SLthres2}) and find the lasing threshold 
\begin{equation}\label{SLthres1}
R^{th}=\frac{\gamma_u\gamma_c({\bf 0})}{\Gamma}.
\end{equation}

We conclude this section by noting
that in the limit of a space independent loss the above results
reduce to the corresponding ones of Sec.\ \ref{subsec:rate}.
\subsubsection{Modified Thomas-Fermi solution}
\label{subsec:modTFsol}
In this section we derive an approximate but analytical expression 
for the stationary state of the modified generalized GPE. 
In the case of the conventional GPE it is the so-called 
Thomas-Fermi (TF) approximation which describes the
steady state of the condensate \cite{TFA}. 
The phase of the stationary solution of the GPE as well as the phase
of the TF approximation is constant.
However, for a position dependent loss it turns out that
the phase of the stationary matter-wave field also depends on the position.
A spatially dependent phase leads to a non-vanishing current of the
condensate. We expect this feature
to be important for the coherence properties of a cw atom laser.
Therefore, we introduce a modified TF solution. 

We start from the time independent form of the modified generalized GPE,
Eq.\ (\ref{time_indep}).
For the present problem a hydrodynamic treatment is more convenient.
We therefore consider the density 
\begin{equation}
\rho({\bf r})\equiv|\psi^{(s)} ({\bf r})|^2 
\end{equation}
and velocity
\begin{equation}
{\bf v}({\bf r})\equiv (\hbar/m)\nabla \phi({\bf r})
\end{equation}
of the condensate. Here $\phi({\bf r})$ is the phase of the mean field
following from the ansatz
\begin{equation}
\psi^{(s)} \left( {\bf r}\right) 
=\sqrt{\rho \left( {\bf r}\right) }e^{i\phi\left( {\bf r}\right) }.
\end{equation}
Indeed, with Eq.\ (\ref{time_indep}) we find the equations
\begin{equation} \label{real_part}
\frac{\hbar^2}{2m}\nabla^2\sqrt{\rho }
=\left[ -\mu +\frac{1}{2}m{\bf v}^2
+V\left( {\bf r}\right) +U_0\rho \right] \sqrt{\rho }
\end{equation}
and
\begin{equation} \label{imag_part}
\frac{{\bf v}}{2}\nabla \sqrt{\rho }
=\left[ -\nabla \cdot {\bf v}-\gamma_c\left( {\bf r}\right) 
+\Gamma N_u\right]\sqrt{ \rho }.
\end{equation}
We make the assumption that the density profile is slowly varying.
Since we are interested in loss shapes where 
the loss at the center of the condensate is uniform, 
these assumptions are reasonable at the center of the condensate. 
However, at the edges of the condensate where the change of the loss is 
large, we expect this assumption to break down.
This is the idea of our modified Thomas-Fermi approximation. 

This assumption allows us to neglect the terms  
$\nabla^2\sqrt{\rho}$ and $\nabla \sqrt{\rho }$ in the two equations.
However, in order to fulfill the second equation we have to retain 
the derivative $\nabla\cdot {\bf v}$ in velocity.
In this approximation we arrive at the approximate expression
\begin{equation} \label{modsqr}
\rho\left( {\bf r}\right) 
\cong \frac{1}{U_0}\left[\mu 
-{\small\frac{1}{2}}m {\bf v}^2\left( {\bf r}\right)
-V\left( {\bf r}\right) \right]
\end{equation}
for the density and the differential equation 
\begin{equation} \label{nablasqr_phi}
\nabla \cdot{\bf v}\left( {\bf r}\right) 
\cong-\gamma_c\left( {\bf r}\right) +\Gamma N_u 
\end{equation}
for the velocity.

Since the density cannot be negative, these expressions are
only valid for that volume ${\cal V}$ of space where 
\begin{equation}\label{MTFradius}
\rho({\bf r})\propto \mu -\frac{1}{2}m
{\bf v}^2\left( {\bf r}\right)-V\left({\bf r}\right)\ge 0.
\end{equation}
The shape of this volume ${\cal V}$ depends on the potential
$V({\bf r})$ and the loss rate $\gamma_c ({\bf r})$ 
via the velocity ${\bf v}({\bf r})$.

Note, that the velocity ${\bf v}$ and consequently the phase $\phi$ 
depends on the rate $\Gamma$, the number of un-condensed 
atoms, and the shape of the loss rate. Indeed, it is only the spatial
dependence of the loss rate that determines
the spatial profile of the velocity ${\bf v}({\bf r})$.
The ``kinetic energy'' of the condensate plays a role similar to 
the trap potential in shaping the density. 

The expressions, Eqs.\ (\ref{modsqr}) and (\ref{nablasqr_phi}),
for the density and velocity are not explicit.
Indeed, the chemical potential $\mu$ is still a free parameter. 
Moreover, the number of un-condensed atoms $N_u^{(s)}$ is coupled to 
Eqs.\ (\ref{modsqr}) and (\ref{nablasqr_phi}) and is given 
by Eq.\ (\ref{ratesolsSL1e}).

We therefore have to solve the various constraints in a self-consistent 
way: The number $N_c^{(s)}$ of condensed atoms reads
\begin{equation}\label{MTFnorm}
N_c^{(s)}=\int_{{\cal V}}\rho({\bf r}) \,d^3r.
\end{equation}
On the other hand this quantity $N_c^{(s)}$
is given by Eq.\ (\ref{ratesolsSL1g}) and reads
\begin{equation}\label{MTFrate}
N_c^{(s)} = \frac{R_u}{\int_{{\cal V}}
\gamma_c({\bf r})\rho_1({\bf r}) \,d^3r}-\frac{\gamma_u}{\Gamma},
\end{equation}
where $\rho_1({\bf r})=[N_c^{(s)}]^{-1}\rho({\bf r})$ is the density 
normalized to unity.
Note, that ${\cal V}$ is defined by the condition that
the density is non-negative, Eq.\ (\ref{MTFradius}).

The number of un-condensed atoms then follows from 
Eq.\ (\ref{ratesolsSL1e}). Hence we have to solve
the three Eqs.\ (\ref{MTFradius}), (\ref{MTFnorm}), and (\ref{MTFrate})
for the three unknowns: the modified Thomas-Fermi volume
${\cal V}$, the chemical potential $\mu$, and the number $N_c^{(s)}$ 
of atoms in the condensate.
\subsubsection{Usual Thomas-Fermi solution}
The most important consequence of our position dependent loss term 
is a non-vanishing velocity and therefore
a spatially dependent phase. How does this compare to the usual
Thomas-Fermi approximation? Here we consider the density 
\begin{equation}\label{usualTFdens}
\rho({\bf r})\cong\frac{\mu-V({\bf r})}{U_0}
\end{equation}
and the velocity
\begin{equation}\label{usualTFvelo}
{\bf v}\equiv 0.
\end{equation}
We first note that this ansatz is in contradiction 
to Eq.\ (\ref{imag_part}). Nevertheless we can use it to investigate 
how the kinetic energy potential influences the density $\rho$. 
Similar to the last section we insert our ansatz into 
Eqs.\ (\ref{MTFradius}) to (\ref{MTFrate}) and find the 
Thomas-Fermi volume ${\cal V}$, the chemical potential $\mu$, 
and the number of atoms $N_c^{(s)}$.
\subsubsection{Exact solution}
\label{subsec:exactsol}
In this section we adapt the numerical methods developed
to find the ground-state solution of the ordinary GPE
\cite{statsols1,statsols2,statsols3} to the present problem. 
This method evolves the wave function for a fixed atom number in 
imaginary time. The solution is normalized to unity after each 
time step. The evolution in imaginary time attenuates
the differences between the arbitrary initial wave function and the
ground-state solution of the ordinary GPE. 

In our model of an atom laser, Eq.\ (\ref{psiSL}), we have generalized 
the GPE by a pump and a loss term. Therefore the number of atoms is 
not fixed but is governed by Eq. (\ref{NeSL}). In order to find the
ground-state solution of our time independent Eqs.\ (\ref{ratesolsSL1g})
and (\ref{time_indep}),
we evolve an arbitrary initial wave function according to our modified 
generalized GPE, Eq.\ (\ref{psiSL}), in imaginary time, using the 
split-operator technique \cite{split1,split2}. After each time step
we normalize the wave function 
to unity. Then we update the number of atoms 
with the help of Eq.\ (\ref{ratesolsSL1g}) and use it
for the next time step. We repeat this procedure 
until the wave function and the atom number has converged
to a stationary value. This method finds the
stationary state of our improved model of an atom laser in a 
self-consistent way. The most important result is that 
the pump and spatial loss gives a space dependent 
phase to the stationary mean-field $\psi^{(s)}({\bf r})$.
\subsection{Results}
As in Sec.\ \ref{sec:model} we now specify the potential 
$V({\bf r})$ to be the one-dimensional harmonic oscillator 
potential, Eq.\ (\ref{harmpot}). Moreover, we study two models 
of a position dependent loss rate: a sum 
\begin{equation}\label{gaussianloss}
\gamma_c(x)=\gamma_c'
\left( \mbox{e}^{-(x+x_0)^2/\sigma^2}
+\mbox{e}^{-(x-x_0)^2/\sigma^2} \right)
\end{equation}
of two Gaussians and a sum 
\begin{equation}\label{spatial_loss}
\gamma_c\left( x\right)=\gamma_c'\left[ 
 \frac{\sigma^2}{\left(x+x_0\right)^2+\sigma^2}
+\frac{\sigma^2}{\left(x-x_0\right)^2+\sigma^2}\right].  
\end{equation}
of two Lorentzians. 

In both cases $\sigma$ is a measure of the width 
of the corresponding distributions
and $\gamma_c'$ is the maximal loss rate when the two curves
have negligible overlap.
In the next section we choose the locations $\pm x_0$ of the maximal loss
such that they sit where the density of the condensate falls off.
\subsubsection{Transient behavior}
First we want to show that the collective excitations which emerged
in the elementary model are damped out due to the presence of the 
space dependence in the loss \cite{damping}. 
For this analysis we use the Gaussian loss term, 
Eq.\ (\ref{gaussianloss}). We adjust $\gamma_c'$ such that we
obtain the same final number of atoms in the condensate
as in the spatially independent case. 
The width of the loss is comparable to a typical scale, 
such as the width of the ground state of the harmonic oscillator.

In Fig.\ \ref{figure1}(a) we compare the time evolution of the 
numbers of condensed and un-condensed atoms for the position 
dependent loss with those of the position independent loss. 
We see that the overall behavior is not that different. 
Therefore our original rate equations of the elementary model
still approximate very well the time evolution of $N_c$ 
and $N_u$ following from the improved model. 

We then show in Fig.\ \ref{figure1}(c) the scaled width of the mean 
field. Indeed, the position dependent loss damps out the collective
excitations, and a steady state is reached. We emphasize that this 
is true for any initial non-vanishing mean field.

In Fig.\ \ref{figure3}(b) we compare and contrast the stationary mean field 
to the numerical solution of the time independent equation.
They coincide with each other. Indeed, for many initial conditions
and set of parameters we have noticed that the mean field obtained by 
evolving the modified generalized GPE over a sufficiently long
enough time such that the transient oscillations are damped out is 
equivalent to the time independent ground-state energy solution of the 
modified generalized GPE. This holds as long as the loss is confined to the
edges of the condensate \cite{genlossrem}. This is the case in the present 
example as shown in Fig.\ \ref{figure3}(b) where
we also display the loss function in arbitrary units. 
The overlap between the stationary mean field and the loss function 
is essential for the number of atoms in the condensate as is apparent
from Eq.\ (\ref{ratesolsSL1g}). 

When we compare Fig.\ \ref{figure3}(a) and (b), we note that
the modulus of the steady-state mean field in the presence of
a loss located at the edges of the condensate is very similar
to the ground-state energy solution of the conventional time 
independent GPE, Eq.\ (\ref{GPeqtimeindep}). However, this is not true 
for the phase of the mean field as already discussed in 
Sec.\ \ref{subsec:statsol}.

A spatially dependent loss term located at the edges of the condensate 
damps out the collective excitations. In our theory of an atom laser this 
plays the role of mode selection, i.e., the ground-state energy solution
survives whereas the excited states, i.e., the collective excitations,
are damped away due to the fact that their overall spread in
position space is greater than that of the ground state.
\subsubsection{Stationary solution}
{\it (a) Modified Thomas-Fermi solution.}
We now turn to the discussion of the modified Thomas-Fermi solution 
derived in Sec.\ \ref{subsec:modTFsol} for a general potential. 
In the present section we restrict this analysis to a one-dimensional 
harmonic oscillator potential, Eq.\ (\ref{harmpot}). Moreover,
we choose 
\end{multicols}
\begin{multicols}{1}
\widetext
\begin{figure}[h] 
\centerline{\epsfig{file=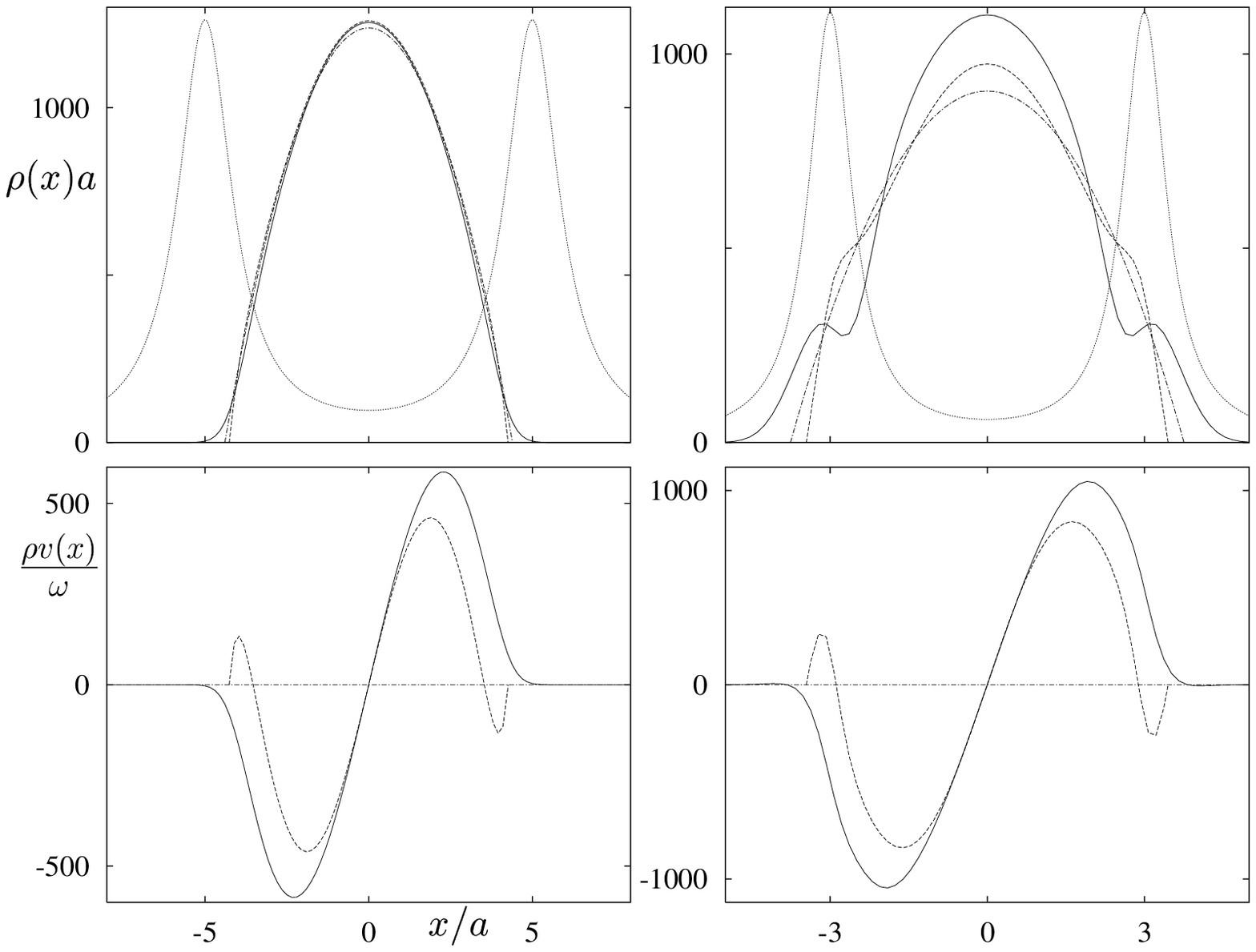,width=17.5cm}}
\end{figure}
\end{multicols}
\begin{multicols}{2}
\narrowtext
\begin{figure}[h] 
\caption{Comparison between the two Thomas-Fermi and the numerical solutions
of the atom laser equations in the presence of space dependent loss.
We display density (upper part) and current (lower part) of the numerical 
(solid line), modified TF (dashed line), and usual TF (dash-dotted line)
solution. On the top we also show the space dependence of the loss term 
$\gamma_c(x)$. Note the position of the peaks, the height is not to scale. 
Here we have used the parameters $R_u/\omega=5\times 10^{3}$, 
$\Gamma/\omega=0.5$, $\gamma_u/\omega=1$, $\gamma_c'/\omega=5$, 
and $U_x/(\hbar\omega a)\protect\cong 0.008$.
In the left column the loss is concentrated at $x=\pm 5a$, whereas
on the right it is at $x=\pm 3a$. Moreover, the widths are different
$\sigma /a=1$ (left) and $\sigma /a=0.5$ (right).}
\label{figure4} 
\end{figure}
\noindent
the Lorentzian loss rate, Eq.\ (\ref{spatial_loss}).
This choice is solely motivated by the fact that we can perform the
resulting integral of the differential equation (\ref{nablasqr_phi}).

We start by first summarizing the 
Eqs.\ (\ref{MTFradius}) to (\ref{MTFrate}) for the chemical potential $\mu$,
the Thomas-Fermi radius $R$, and the number of condensed atoms $N_g^{(s)}$,
\begin{eqnarray}
\label{MTFradius1D}
\mu -\frac{1}{2}m\left[ 
v\left(R\right)\right]^2-\frac{1}{2}m \omega^2 R^2 &=& 0,\\
\label{MTFnorm1D}
N_c^{(s)}-\int_{-R}^R\rho(x) \,dx&=&0,\\
\label{MTFrate1D}
N_c^{(s)} - \frac{R_u}{\int_{-R}^R
\gamma_c(x)\rho_1(x) \,dx}-\frac{\gamma_u}{\Gamma}&=&0,
\end{eqnarray}
determining the modified Thomas-Fermi solution for the
one-dimensional harmonic trap and loss. 
From Eq.\ (\ref{modsqr}) we obtain the one-dimensional density
\begin{equation} \label{density1D}
\rho\left( x\right) 
= \frac{1}{U_x}\left[\mu 
-{\small\frac{1}{2}}m v^2\left(x\right)
-\frac{1}{2}m \omega^2 x^2\right].
\end{equation}
We find the velocity $v$
when we substitute the Lorentzian loss rate, Eq.\ (\ref{spatial_loss}),
into Eq.\ (\ref{nablasqr_phi}) and integrate which yields
\begin{eqnarray}\label{velocity1D}
v\left( x\right) 
&=& \Gamma N_u x-\gamma_c'\sigma \left[ \arctan \left( 
\frac{x+x_0}{\sigma}\right) \right. \nonumber \\
&&\left. +\arctan \left( \frac{x-x_0}{\sigma }\right) \right].
\end{eqnarray}
Here we have set $v(0)=0$ 
in order to preserve the symmetry of the solution of the mean field, 
Eq.\ (\ref{density1D}), 
$\rho(x)=\rho(-x)$.
The latter holds because the harmonic oscillator potential, the spatially
dependent loss, and the generalized GPE show this symmetry. 

{\it (b) Usual Thomas-Fermi solution.}
In order to get a feeling how in our one-dimensional model the
density $\rho$ is influenced by the velocity $v$ we also discuss
the usual TF solution. For the one-dimensional harmonic trap
the density, Eq.\ (\ref{usualTFdens}), reads
\begin{equation}\label{usualTFdens1D}
\rho(x)\cong\frac{1}{U_x}\left[\mu-m \frac{1}{2}\omega^2 x^2\right],
\end{equation}
and the velocity, Eq.\ (\ref{usualTFvelo}), reads 
\begin{equation}\label{usualTFvelo1D}
v(x)\equiv 0. 
\end{equation}
When we use this usual TF approximation, we easily find
from Eq.\ (\ref{MTFradius1D}) with $v\equiv0$ the
TF radius $R=\sqrt{2\mu/m\omega^2}$.
We then use Eq.\ (\ref{MTFnorm1D}) to calculate the number of 
atoms in the condensate as a function of the chemical potential $\mu$.
Inverting this equation we can express the chemical potential 
$\mu=[3 N_c^{(s)} U_x\sqrt{m}\omega/(4\sqrt{2})]^{2/3}$
as a function of the atom number $N_c^{(s)}$.
Finally, we use Eq.\ (\ref{MTFrate1D}) to determine the
number $N_c^{(s)}$ of condensed atoms, the only unknown quantity.
We emphasize that only this last step has to be done numerically.

{\it (c) Discussion.}
Figure \ref{figure4} shows the modified TF, the usual TF and the
fully numerical solution.
We have chosen two different shapes of the Lorentzian loss curve:
In the right column the width of the individual Lorentzians is
half the size of the one on the left column.
Moreover, we note that their location is different.
We depict the modified TF solution by 
dashed curves, the usual TF solution by dash-dotted ones, and
the numerical solution by solid lines. Again we show in arbitrary units
the shape of the loss $\gamma_c\left( x\right) $ by the dotted curve. 

The density is shown in the upper part of Fig.\ \ref{figure4}.
We see that for the parameters of the left column 
both TF approximations for the density work quite well.
The modified TF solution approximates the
center better than the usual TF solution. 
However, the radius derived from the usual TF approximation is a 
better estimate of the edge of the condensate than the modified TF radius.
For the set of parameters used in the right column of 
Fig.\ \ref{figure4} both TF approximations for the density do not work
that well anymore. The reason for this break-down is that
the condensate reaches too far
into the loss region where the density varies strongly
and the derivatives neglected in the derivation of Eqs.\ (\ref{modsqr}) 
and (\ref{nablasqr_phi}) become important. Nevertheless
our modified TF solution shows at least qualitatively the same behavior
as the fully numerical solution.
Surprisingly, at regions towards the peaks of the loss where the loss rate
is at its highest, the density is first decreased and then increased
in the modified TF over the usual TF 
solution. This is understandable from Eq.\ (\ref{modsqr})
when we consider in the lower part of Fig.\ \ref{figure4} the current 
$j(x)=\rho(x) v(x)$. We note that at these regions first
the current which is proportional to the velocity 
is at its highest and then decreases.

We see that for the current, i.e.\ the velocity, the modified 
TF solution is a very good approximation around 
\begin{figure}[h] 
\centerline{\epsfig{file=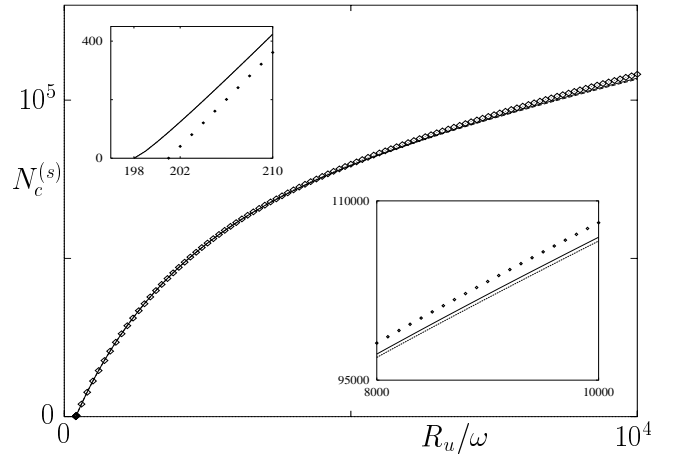,width=8.5cm}}  
\caption{
Comparison between analytical, but approximate, and numerical, 
but exact solutions of the atom laser equations. For this 
comparison we analyze the threshold behavior of the atom laser: 
Condensate population $N_c^{(s)}$ in steady-state as a function 
of the pump strength $R_u$. Over a wide range of pump strengths 
both approximate solutions, the modified (solid line) and the 
usual (dashed line) TF solution accurately describe the fully 
numerical solution (diamonds). In the region close to threshold 
the two approximate solutions, which are almost identical, clearly 
deviate from the fully numerical solution as shown in the inset 
in the upper left corner. The other inset magnifies a region far 
above threshold where the modified TF solution is a better 
approximation than the usual TF solution. The parameters used
are: $\Gamma/\omega=10^{-3}$, $\gamma_u/\omega=10$, 
$\gamma_c'/\omega=1$, $\sigma /a=1$ and $x_0/a=10$, 
and $U_x/(\hbar\omega a)\protect\cong 0.008$.
} 
\label{figure5} 
\end{figure} 
\noindent
the central region. 
It gives a good estimate for the overall behavior of the current 
(velocity) for both sets of parameters. Around the central region of 
the condensate the current is approximately a linear function of the 
position, illustrating the flow of 
atoms to the ends where they are predominately lost from the peaks. 
At the tails of the condensate the modified TF solution even predicts
a velocity changing the direction. This is in contrast to the fully 
numerical solution. The usual TF solution has zero velocity by default. 

We solve numerically the Eqs.\ (\ref{MTFradius1D}) to (\ref{MTFrate1D}) 
for the modified Thomas-Fermi solution to find the condensate
population for various pump strengths and display this as the
solid curve in Fig.\ \ref{figure5}. The dashed curve corresponds to
the usual TF solution.
We display the fully numerical solution with the help of diamonds.

Both approximate analytical curves agree quite well with
the fully numerical solution over a wide region of different pump strengths.
However, at pump rates just above threshold 
the agreement is not that good which is shown in the upper left inset
of Fig.\ \ref{figure5} .
Both the usual TF and the modified TF curves lie on top of each other and
over-estimate the numerical result.
The fully numerical solution crosses the horizontal axis at 
$R_u/\omega\cong 201$ whereas the two TF solutions cross at
$R_u/\omega\cong 198$. 
Close to threshold, the spatial extent of the two approximate wave 
functions is becoming smaller with decreasing atom number. They are
predominately located around the relatively flat loss region and 
the losses appear to be position independent. Therefore the laser 
threshold for the two approximate solutions is predicted by 
Eq.\ (\ref{SLthres1}). In contrast, the shape of 
the numerically calculated mean field can vary, and is 
in fact of Gaussian form close to threshold. 
Hence our approximation of the threshold, Eq.\ (\ref{SLthres2}),
agrees rather remarkably with the numerically calculated threshold.
The deviation of the approximate threshold,
Eq.\ (\ref{SLthres1}), reflects the fact that even close to threshold
the loss function does not appear spatially constant to the mean field
for the parameters of Fig.\ \ref{figure5}.
When we recall that the TF approximation is not good at low atom numbers
it is rather surprising to see such small differences
between the TF solutions and the numerical solution.

The lower right inset of Fig.\ \ref{figure5}
zooms in on a region of higher pump strengths far above threshold 
where the spatial structure
of the loss plays a role. Here the modified TF solution approximates
the fully numerical solution (diamonds) better than
the usual TF solution. This is understandable since the modified TF
solution takes into account pump and spatially dependent loss
by allowing for a spatially dependent phase, i.e.\ velocity.
\section{Conclusions}
\label{sec:concludere}
In summary, we have constructed a theory of an atom laser that 
is analogous to semi-classical laser theory. The
matter-wave equation is a generalized Gross-Pitaevskii equation
with additional loss and gain terms. We derive the lasing threshold and
describe the build-up of the coherent mean field of a condensate. 

The elementary model uses a spatially homogeneous
loss. Here we find un-damped collective excitations. 
Therefore the final mean field depends
on the initial mean field: The known
stationary ground state of the GPE which is the desired lasing mode 
cannot be reached in general. 

The improved model has 
a natural mode selection built in by a space dependent loss. 
In this way we achieve the desired single lasing mode.

We have derived a modified Thomas-Fermi solution 
for the steady-state mean field.
This solution takes into account the effects
of a pump term and a position dependent loss term. 
In contrast to a constant phase of the usual Thomas-Fermi solution,
the modified Thomas-Fermi solution has a spatially dependent
phase, i.e.\ velocity, due to the permanent flow of atoms in and out 
of the condensate. The modified Thomas-Fermi solution is a good 
approximation in regions where the loss shape is slowly varying
and for sufficiently large atom numbers. 

We emphasize that our model of an atom laser is very simple and rather 
general.
Therefore, we can apply it to different experimental configurations of 
cw atom lasers \cite{konstanz}, or current experiments 
\cite{BEC1,BEC2,BEC3,BEC4,BEC5,BEC6,BEC7,BEC8,BEC9,BEC10,BEC11,BEC12} provided
the evaporative cooling process (boson amplification) and the 
loading of the trap is run continuously, and an output coupling 
mechanism is applied.
\acknowledgments 
We thank Eric Bolda, Michael Fleischhauer, Murray Olsen, Karl Riedel, 
and Janne Ruostekoski for stimulating and 
valuable discussions. Two of us (B.K.\ and K.V.) acknowledge very
gratefully the warm hospitality given to them at their stay
with the quantum optics group of the department of physics at 
the University of Auckland. This work was
supported by the Deutsche Forschungsgemeinschaft, the University 
of Auckland Research Committee and the Marsden
Fund of the Royal Society of New Zealand.
\begin{appendix} 
\section*{Collective excitations in one dimension} 
Collective excitations are usually discussed in three  
dimensions \cite{excitations}.  
Within the framework of the Thomas-Fermi approximation, 
Stringari \cite{coll2} has calculated analytically the excitation 
spectrum of a condensate in a three-dimensional isotropic harmonic 
trap. Since our numerical solution of Eq.\ (\ref{psi}) is done for 
a one-dimensional harmonic trap, we apply Stringari's method to 
a one-dimensional harmonic trap of frequency $\omega$. We substitute
\begin{equation} 
\psi(x,t) = \sqrt{\rho(x,t)} e^{i\phi(x,t)} 
\end{equation} 
into Eq.\ (\ref{psi}) and obtain after some algebra the hydrodynamic 
equations 
\begin{equation} 
\frac{\partial \rho}{\partial t} 
+ \frac{\partial}{\partial x} \left(\rho v\right) 
+ \left(\gamma_c - \Gamma N_u \right) \rho = 0 
\end{equation} 
and 
\begin{eqnarray} 
m \frac{\partial v}{\partial t}  
+ \frac{\partial}{\partial x} \left( \frac{m}{2} v^2  
+ V + U_x \rho -\mu \right)&& \nonumber\\ 
- \frac{\hbar^2}{2m} \frac{\partial}{\partial x} \frac{1}{\sqrt{\rho}} 
\frac{\partial^2}{\partial x^2} \sqrt{\rho} &=& 0 
\label{hydro-v} 
\end{eqnarray} 
for the ``density''  
\begin{equation} 
\rho(x,t) = \psi^{\ast}(x,t) \psi(x,t) 
\end{equation} 
and the ``velocity''  
\begin{equation} 
v(x,t) = \frac{\hbar}{m} \frac{\partial \phi(x,t)}{\partial x} . 
\end{equation} 
Note, that we have also introduced the chemical potential $\mu$ which is  
space independent. 
 
We now consider small deviations 
\begin{eqnarray} 
\delta \rho &\equiv& \rho - \rho^{(s)},  \nonumber\\ 
\delta v &\equiv& v - v^{(s)} = v, \nonumber\\ 
n_u &\equiv& N_u- N_u^{(s)} =N_u- \gamma_c/\Gamma
\end{eqnarray} 
of $\rho$, $v$, and $N_u$ from  
their stationary values $\rho^{(s)}$, $v^{(s)}\equiv 0$, and 
$N_u^{(s)} = \gamma_c/\Gamma$.
Furthermore, we make the Thomas-Fermi approximation, that is, we 
neglect the last term in Eq.\ (\ref{hydro-v}) and approximate the 
stationary solution $\rho^{(s)}(x)$ by the Thomas-Fermi solution \cite{TFA}
\begin{equation} 
\rho^{(s)}(x) \cong \frac{\mu - V(x)}{U_x},
\end{equation} 
where the chemical potential $\mu$ is defined by the normalization
integral. We then arrive at the linearized equations 
\begin{equation} 
\frac{\partial}{\partial t} \delta \rho  
+ \frac{\partial}{\partial x} \left( \rho^{(s)} v \right)  
- \Gamma \rho^{(s)} n_u = 0 
\end{equation} 
and 
\begin{equation} 
m \frac{\partial v}{\partial t}  
+ U_x \frac{\partial}{\partial x}\delta \rho = 0 . 
\end{equation} 
We combine these two equations to eliminate $v$ and arrive at 
\begin{equation} 
\frac{\partial^2}{\partial t^2} \delta \rho 
- \frac{U_x}{m}\frac{\partial}{\partial x} 
\left( \rho^{(s)}  
\frac{\partial}{\partial x} \delta \rho \right)  
= \Gamma \rho^{(s)} \dot{n}_u . 
\label{xxx} 
\end{equation} 
For large times, when $N_u(t)$ and $N_c(t)$ have already 
reached their stationary value, we can neglect the inhomogeneous term 
$\Gamma \rho^{(s)} \dot{n}_u$. 
 
Equation (\ref{xxx}) cannot be solved without knowledge of 
the potential $V(x)$. We restrict ourselves to the case of a harmonic trap, 
that is,  
\begin{equation} 
V(x) = \frac{1}{2}m\omega^2 x^2.  
\end{equation} 
In order to solve Eq.\ (\ref{xxx}), 
we introduce the scaled variable $\xi = x/R$, where 
$R=\sqrt{2\mu/m\omega^2}$ 
is the Thomas-Fermi radius of the condensate. Using the ansatz 
\begin{equation} 
\delta \rho(x,t) = A \sin(\Omega t + \varphi) y(x/R) 
\end{equation} 
we obtain an ordinary differential equation for $y(\xi)$ which reads 
\begin{equation} 
\frac{d}{d\xi} \left[(1-\xi^2)\frac{dy(\xi)}{d\xi}\right]  
+ \frac{2\Omega^2}{\omega^2} y(\xi) = 0. 
\label{legendre} 
\end{equation} 
This is the differential equation of Legendre functions 
which in general only has solutions that are singular at  
$\xi=\pm 1$ \cite{courant}, that is at $x=\pm R$.  
The only exceptions are 
\begin{equation} 
\frac{2\Omega_n^2}{\omega^2} = n(n+1), 
\label{freq} 
\end{equation} 
where $n$ is an integer. In this case the well-known 
Legendre polynomials \cite{courant} 
\begin{equation} 
P_n(\xi) = \frac{1}{2^n n!} \frac{d^n}{d\xi^n}\left(\xi^2-1\right)^n 
\end{equation} 
solve Eq.\ (\ref{legendre}). Furthermore, they fulfill 
the orthogonality relation 
\begin{equation} 
\int\limits_{-1}^{+1} P_n(\xi) P_m(\xi) \,d\xi = 0 
\quad \mbox{for} \quad n \ne m . 
\label{ortho} 
\end{equation} 
The frequencies of the elementary excitations are therefore given by 
Eq.\ (\ref{freq}). The solution of the homogeneous part of  
Eq.\ (\ref{xxx}) reads 
\begin{equation}\label{soldev} 
\delta \rho(x,t)  
= \sum_{n=1}^{\infty} A_n \sin(\Omega_n t + \varphi_n) P_n(x/R) , 
\end{equation} 
where $A_n$ and $\varphi_n$ follow from the initial deviation 
from the stationary solution. 

Equation (\ref{soldev}) shows that the excitations do not decay, 
even in the presence of pump and loss terms in our generalized 
Gross-Pitaevskii equation, Eq.\ (\ref{psi1D}). However, these 
excitations do not grow but
rather oscillate. Our numerical solution of Eq.\ (\ref{psi1D}) 
in Sec.\ \ref{subsec:num} confirms this result.

In Fig.\ \ref{figure1}(b) 
we do not find all frequencies allowed by Eq.\ (\ref{freq}) 
since in our numerical  
solution of Eqs.\ (\ref{psi1D}) and (\ref{Ne1D}) we  
start with a symmetric initial condition $\psi(x,t=0)$.
Because Eqs.\ (\ref{psi1D}) and (\ref{Ne1D}) 
do not destroy this symmetry, $\rho(x,t)$  
as well as $\delta\rho(x,t)$, the deviation from the (symmetric)  
stationary solution of the generalized GPE, will always be symmetric. 
We therefore can only find excitations which correspond to Legendre 
polynomials of even order if we start with a symmetric $\psi(x,t=0)$. 
The corresponding frequencies are 
\begin{equation} 
\Omega_{2n} = \omega \sqrt{n(2n+1)} . 
\end{equation} 
This is also true for an anti-symmetric initial wave function
since the corresponding density is symmetric.

Two other facts are worth mentioning: The excitations discussed 
above do not change the number of atoms. Using $P_0(\xi) = 1$ 
we find with the help of the orthogonality relation Eq.\ (\ref{ortho}) 
\begin{equation} 
\int \delta \rho(x,t) \,dx = 0 .  
\end{equation} 
The second moment
\begin{equation} 
\overline{x^2(t)} = \frac{\int x^2 |\psi(x,t)|^2 \,dx}
                         {\int |\psi(x,t)|^2 \,dx}
\end{equation} 
can only oscillate with the frequency $\Omega_2=\sqrt{3}\,\omega$. 
This follows from the relation 
\begin{equation} 
\xi^2 = \frac{2}{3} P_2(\xi) + \frac{1}{3} P_0(\xi) 
\end{equation} 
together with the orthogonality relation, Eq.\ (\ref{ortho}). 
A generalization is the statement that  
\begin{equation} 
\overline{P_k(x/R)} = \frac{\int P_k(x/R) |\psi(x,t)|^2 \,dx}
                           {\int |\psi(x,t)|^2 \,dx}
\end{equation} 
only shows oscillations with frequency $\Omega_k$. 
\end{appendix} 
\end{multicols}

\begin{references}
\bibitem{dipdip}
Due to the dipole-dipole interaction of the atoms it is sometimes
more convenient to use instead of the mode functions of the empty
resonator the solutions of the time independent Gross-Pitaevskii
equation.
\bibitem{discnumrem}
We are aware of the fact that the number of atoms is discrete. However,
similar to the semi-classical theory of the optical laser where the field
intensity is continuous, the definition of $N_c$, Eq.\ (\ref{DefNg}), 
allows any positive value for the number of atoms.
\bibitem{noteqmexp}
Note, that these moments are not the quantum mechanical expectation
values $\langle\hat{x}\rangle$ and $\langle\hat{x}^2\rangle$ of 
the position operator 
$\hat{x}(t)\equiv\int\hat{\psi}^\dagger(x,t)\,x\,\hat{\psi}(x,t)\,dx$.
The quantities $\langle\hat{x}\rangle$ and $\langle\hat{x}^2\rangle$
do not need the normalization
introduced in Eqs.\ (\ref{firstmom}) and (\ref{secondmom}).
Moreover $\langle\hat{x}^2\rangle$ 
involves the operator $\hat{\psi}$ in a quartic way.
\bibitem{reviv}  
For a three-dimensional anisotropic trap these oscillations can 
collapse and revive again due to mode beating. See for example 
Refs.\ \cite{reviv1,reviv2,reviv3,reviv4}.
\bibitem{damping}  
It has been shown that collective 
excitations of a condensate can be damped due to several mechanisms
\cite{damping1,damping2,damping3,damping4,damping5}.
\bibitem{genlossrem} 
The fact that the ground state solution survives is because of the 
loss shape. For example, a loss shape which is located at the center 
preferably damps away the ground state solution. Therefore an excited 
state survives the mode competition.
\end{references}
\end{document}